# Task-driven assessment of experimental designs in diffusion MRI: a computational framework

Sean C. Epstein[1], Timothy J.P. Bray[2], Margaret A. Hall-Craggs[2], Hui Zhang[1]

[1]Department of Computer Science & Centre for Medical Image Computing, University College London, United Kingdom.

[2]Centre for Medical Imaging, University College London, London, United Kingdom.

## Word count

Abstract: 259

Main body (not including references & figure captions): 3192

Figure captions: 294

## Abbreviations

In order of appearance in the text:

| | |
|---|---|
| CED | Computational experimental design |
| ROC | Receiver operating characteristic |
| AUC | Area under the curve |
| dMRI | Diffusion MRI |
| AIC | Akaike information criterion |
| BIC | Bayesian information criterion |
| SpA | Spondyloarthritis |
| SIJ | Sacroiliac joint |
| IVIM | Intravoxel incoherent motion |
| ADC | Apparent diffusion coefficient |
| bcNLLS | Bound-constrained non-linear least squares |
| sNLLS | Segmented non-linear least squares |
| DTI | Diffusion tensor imaging |

# ABSTRACT

This paper proposes a task-driven computational framework for assessing diffusion MRI experimental designs which, rather than relying on parameter-estimation metrics, directly measures quantitative task performance. Traditional computational experimental design (CED) methods may be ill-suited to experimental tasks, such as clinical classification, where outcome does not depend on parameter-estimation accuracy or precision alone. Current assessment metrics evaluate experiments' ability to faithfully recover microstructural parameters rather than their task performance. The method we propose addresses this shortcoming.

For a given MRI experimental design (protocol, parameter-estimation method, model, etc.), experiments are simulated start-to-finish and task performance is computed from receiver operating characteristic (ROC) curves and associated summary metrics (e.g. area under the curve (AUC)). Two experiments were performed: first, a validation of the pipeline's task performance predictions against clinical results, comparing in-silico predictions to real-world ROC/AUC; and second, a demonstration of the pipeline's advantages over traditional CED approaches, using two simulated clinical classification tasks. Comparison with clinical datasets validates our method's predictions of (a) the qualitative form of ROC curves, (b) the relative task performance of different experimental designs, and (c) the absolute performance (AUC) of each experimental design. Furthermore, we show that our method outperforms traditional task-agnostic assessment methods, enabling improved, more useful experimental design.

Our pipeline produces accurate, quantitative predictions of real-world task performance. Compared to current approaches, such task-driven assessment is more likely to identify experimental designs that perform well in practice. Our method is not limited to diffusion MRI; the pipeline generalises to any task-based quantitative MRI application, and provides the foundation for developing future task-driven end-to end CED frameworks.

# 1 – INTRODUCTION

When planning quantitative diffusion MRI (dMRI) experiments, investigators can select from a range of acquisition and analysis parameters, such as diffusion model, acquisition protocol, and parameter-estimation method. In this context, experimental design refers to the process by which these choices are (i) proposed, (ii) assessed, and (iii) optimised. *Computational* experimental design (CED) describes in-silico approaches to this process, by which experiments are improved without the need for costly clinical datasets[1–10].

Assessment of experimental choices is central to CED, and the choice of assessment metric is critical in determining CED outcomes. In current approaches, these metrics evaluate an experiment's ability to faithfully recover microstructural parameters of interest: bias-minimising information-theory metrics (e.g. AIC, BIC) are used to select microstructural models[11–13]; bias and/or variance-minimising metrics are used to select both acquisition protocols[4,9,10,14–27] and parameter-estimation methods[28–31].

We argue that such assessment may be ill-suited to contexts involving tasks, such as classifying subjects as healthy or diseased, where outcome does not depend solely on parameter accuracy or precision. In such settings, the ultimate metric of interest is *effect size*, not microstructural fidelity. Indeed, gold-standard non-computational assessment techniques are task-driven: experimental designs are assessed on their ability to correctly classify subjects with known diagnoses[32,33]. In contrast, in computational settings, the use of task-agnostic parameter-estimation metrics may lead to unreliable assessment of task performance.

This observation, if verified, impacts the entire experimental design pipeline: high-bias experimental choices (e.g. simplistic dMRI models), which are currently rejected outright, may increase task performance via low parameter-estimation variance; model-fitting technique performance, which may vary on a parameter-by-parameter basis, should be selected on a task-by-task basis.

In this work we propose a pipeline for assessing experimental design that both verifies and addresses these limitations. Given a set of experimental design settings, our pipeline simulates data acquisition, analysis, and task evaluation; task performance is then calculated using receiver operating characteristic (ROC) curves. These curves, which plot classification power as a function of discrimination threshold, can be analysed to produce summary task-performance metrics such as area under the curve (AUC). These metrics can be used to (i) select task-optimal experimental designs from a range of candidate options or (ii) directly underpin computational optimisation. In so doing, we mimic gold-standard non-computational task-driven assessment (experimental repetition), without the associated cost of real-world data acquisition.

This paper is structured as follows: Section 2 details our proposed CED assessment method; Sections 3 and 4 validate our method with data from clinical studies and compare it to traditional task-agnostic approaches based on parameter-estimation metrics. Section 5 analyses the advantages and limitations of our proposed pipeline and discusses how it provides the missing ingredient for developing future task-driven CED frameworks.

## 2 – THEORY

This section presents an overview of our assessment pipeline. Section 2.1 details its design and implementation, and Section 2.2 outlines the advantages it offers over current approaches.

### 2.1 – Proposed pipeline

Given a quantitative task and set of experimental design choices, our pipeline predicts task performance using quantitative summary metrics. For classification tasks, these metrics are typically ROC curves and their associated AUC. The pipeline naturally accommodates other metrics that may be more appropriate for a given application.

The pipeline mimics, in-silico, empirical task-driven assessment: gathering real-world data and measuring sensitivity and specificity. Its structure is shown in Figure 1; it takes three inputs: a quantitative task (I1), characterisation of relevant tissue(s) (I2), and a candidate experimental design (I3). The pipeline combines these inputs to predict associated task performance. It simulates complete dMRI experiments: noisy dMRI data is synthesised (P1), dMRI model parameters are estimated (P2), and task performance is evaluated (P3).

### 2.1.1 – Inputs

*I1:* A quantitative parameter-driven task, characterised by an operational definition and quantitative performance metric.

*Example*: Classification of tissue as either healthy or diseased, based on dMRI parameter estimates: a parameter is chosen, a threshold is set, and a tissue is diagnosed based on its estimated model parameter(s). Task performance is measured by the AUC of an ROC curve computed across a patient population (by sweeping the parameter threshold across all values).

*I2:* Characterisation of the tissues involved in the quantitative task (I1).

*Example:* The 'ground-truth' dMRI model that has been deemed to most-faithfully represent the underlying tissue; associated empirical parameter values (e.g. mean + standard deviation) of each tissue type.

*I3:* Candidate experimental design choices.

*Example:* dMRI acquisition scheme (b-values), dMRI model (may differ from the 'ground-truth' model in I2, see §2.2), parameter(s) selected for classification, parameter-estimation method, signal-to-noise ratio (SNR) (from TE, TR, number of repetitions, average size of region of interest (ROI)), etc.

### 2.1.2 – Simulation

*P1:* Data synthesis.

*Example:* A large number, sufficient to reduce sampling errors (e.g. 10,000), of dMRI signals are synthesised according to I2 for each tissue type (healthy, diseased), using the designated model, drawing from the associated parameter distribution. Each dMRI signal is sampled and corrupted with Rician noise as detailed in I3.

*P2:* Parameter estimation.
*Example:* Each sampled noisy signal is analysed using the specified fitting method (I3) to generate parameter estimates.

*P3:* Task evaluation.
*Example:* The task (I1) is evaluated for the complete set of parameter estimates (P2), and an ROC curve is generated.

### 2.1.3 – Output

*O1:* Task performance.
*Example:* AUC computed from ROC curves output as summary metric of I1 task performance associated with I3 experimental design and I2 tissue.

In this way, for a given I1-I3, the pipeline outputs a prediction of associated task performance. If a wide range of competing candidate experimental settings are assessed, the pipeline's task performance predictions can be used to select the optimal experimental design.

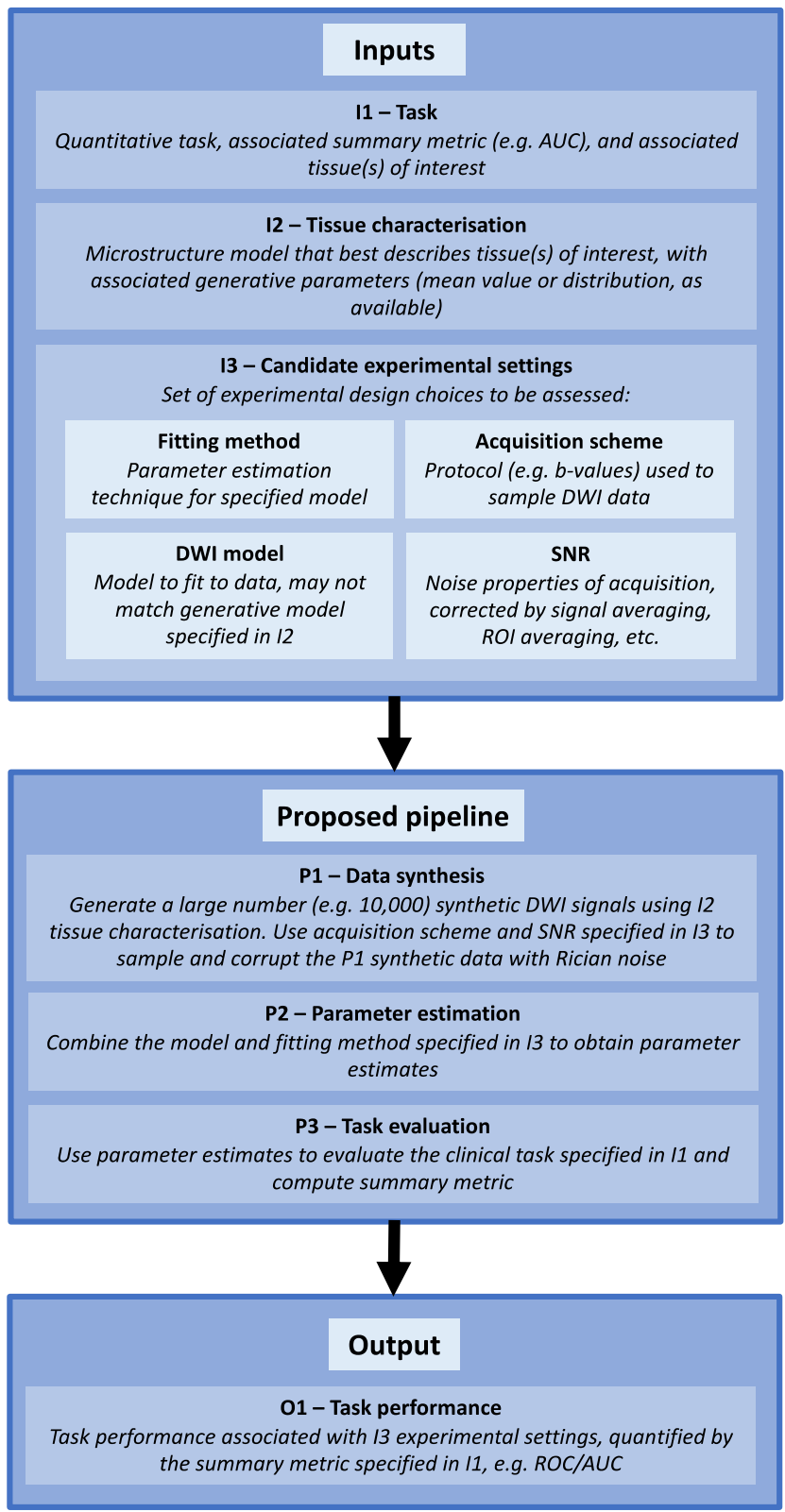

*Figure 1 – Graphical overview of proposed CED assessment pipeline*

**2.2 – Advantages over current approaches**

Our pipeline offers three broad advantages over current CED assessment.

Firstly, it directly assesses task performance. This is a marked difference from existing parameter estimation metrics, such as bias and variance, which give indirect measures of effect-size. This direct assessment enables the selection of experiments which reliably maximise task performance.

Secondly, in contrast to traditional approaches, our pipeline does not automatically reject high-bias, low-variance signal models. These models, which describe tissues less accurately than 'ground-truth' models (I2), may nonetheless increase task performance, especially in settings where effect size depends more strongly on precision and repeatability than microstructural fidelity. Our pipeline enables the assessment and selection of these high-bias yet potentially task-optimal models.

Thirdly, the proposed method assesses parameter-estimation methods in an explicitly task-specific, rather than tissue-specific, manner. Within a single tissue type, a range of tasks may be required, with each task depending on different parameter estimates (see §3.3 for an example). Since the relative performance of different fitting algorithms may vary on a parameter-by-parameter basis[32,34,35], the optimal fitting method may vary *between tasks within a single tissue type*. This kind of task-specific assessment is inaccessible to current CED approaches.

# 3 – METHODS

Two experiments were performed: (E1) a validation of our computational predictions of real-world task performance, and (E2) a demonstration of the benefits afforded by these predictions.

**3.1 – Clinical context**

We used spondyloarthritis (SpA), an inflammatory disease which affects the bone and joints, as a clinical example with which to validate and demonstrate the value of our approach. This choice was made because not only is dMRI increasingly being used as a tool for assessing inflammation in SpA[36–38], but there is an established literature providing our pipeline's required inputs (I1-I3) and associated empirical measures of task performance, against which we can validate our pipeline's outputs (O1).

SpA is characterised by a range of abnormalities which include subchondral bone marrow oedema near the sacroiliac joint (SIJ) margins[39]. These inflammatory lesions, which can be subtyped as either 'active' or 'chronic', have been found to be well-described by the intravoxel incoherent motion (IVIM) dMRI model[40], with IVIM parameter estimates being sensitive to changes in pathology[38,41].

**3.2 – Validation (E1)**

We identified two clinical datasets for which classification task performance was either published or could be computed: an in-house 28-patient dataset ("Bray")[41] split across two SpA subtypes and analysed for one task, and an external 41-patient dataset ("Zhao")[38] split across three SpA subtypes, which was not available for primary analysis but which reported ROC/AUC for three tasks.

For each dataset, our pipeline mirrored real-world experimental design choices (Table 1): tasks involved classifying mean ROI parameter estimates as belonging to one of two SpA subtypes, task performance was assessed with ROC curves and associated AUCs, and data was synthesised from the IVIM model:

$$\frac{S(b)}{S_0} = f e^{-b(D_{fast}+D_{slow})} + (1-f) e^{-bD_{slow}} \quad (1)$$

where $S(b)$ is the MRI signal at diffusion weighting $b$, $S_0$ is the signal at $b = 0$, $f$ is the perfusion fraction, $D_{fast}$ is the pseudo-diffusivity of perfusing water, and $D_{slow}$ is the diffusivity of non-perfusing water.

IVIM tissue parameters were drawn from Gaussian distributions representing SpA lesions relevant to each classification task, and Rician noise was added at SNRs commensurate with each real-world acquisition:

$$S_{noisy}(b) = \sqrt{\mathcal{N}(S_{noisefree}(b), \sigma^2)^2 + \mathcal{N}(0, \sigma^2)^2} \, ; \, SNR = S_0/\sigma \qquad (2)$$

Where $S_{noisefree}$ is the noise-free IVIM signal, $S_0$ is the noise-free signal at $b = 0$, and $\mathcal{N}(\mu, \sigma^2)$ is a Gaussian distribution of mean $\mu$ and standard deviation $\sigma$.

Data was sampled at b-values matching those used in the associated clinical datasets. The signal was normalised by $S_0$ and IVIM fitting was performed with either segmented NLLS (sNLLS[35]) or bound-constrained NLLS (bcNLLS[35]): $f \in [0,1]$, $D_{fast} \in [0,500]$ $10^{-3}$ mm²/s, $D_{slow} \in [0,10]$ $10^{-3}$ mm²/s; fitting was seeded with mean parameter values across tissues within each dataset to mimic real-world experiments in which individual patient classification is not known. In the case of two-step sNLLS fitting, an appropriate choice of $b_{threshold}$ should be made based on expected tissue properties ($f$, $D_{fast}$); in this instance, we replicated Zhao's choice of $b_{threshold}$ = 200 s/mm².

In E1.4, a second signal model (ADC) was fit to the synthetic IVIM data to mimic Bray's clinical investigations. The ADC model takes the form:

$$\frac{S(b)}{S_0} = e^{-b(ADC)} \qquad (3)$$

Where $S(b)$ is the MRI signal at diffusion weighting $b$, $S_0$ is the signal at $b = 0$, and $ADC$ is the apparent diffusion coefficient.

This model was fit on the log-transformed signal using weighted least-squares (WLS) linear regression[42], a single-shot non-iterative method for obtaining maximum likelihood estimates.

Our in-silico ROC/AUC predictions were validated across multiple microstructural models (IVIM, ADC) and model parameters ($D_{fast}$, $D_{slow}$, $f$, ADC) in two ways. Firstly, we simulated large patient populations (1000 x clinical dataset size) to obtain numerically robust task performance predictions. Secondly, to quantify agreement between these results and the smaller clinical datasets, we repeatedly sub-sampled the simulated data, each time matching real-world patient numbers. The resulting distribution of sub-sampled task performance metrics was compared to the empirically observed AUC.

| Dataset | Task | Tissue | Generative model | $f$ | $D_{slow}$ ($10^{-3}\ mm^2/s$) | $D_{fast}$ ($10^{-3}\ mm^2/s$) | Effective SNR | Sampling ($s/mm^2$) | Model & fitting method |
|---|---|---|---|---|---|---|---|---|---|
| Zhao | E1.1 | Chronic | Intravoxel incoherent motion (IVIM) | 0.12 ± 0.02 | 0.35 ± 0.11 | 124.7 ± 13.7 | 150.6 | 0, 10, 20, 30, 50, 80, 100, 200, 400, 800 | IVIM: sNLLS |
| | | Healthy | | 0.09 ± 0.02 | 0.34 ± 0.09 | 122.7 ± 18.3 | | | |
| | E1.2 | Active | | 0.12 ± 0.03 | 0.99 ± 0.39 | 123.9 ± 19.9 | | | |
| | | Chronic | | 0.12 ± 0.02 | 0.35 ± 0.11 | 124.7 ± 13.68 | | | |
| | E1.3 | Active | | 0.12 ± 0.03 | 0.99 ± 0.39 | 123.9 ± 19.9 | | | |
| | | Healthy | | 0.09 ± 0.02 | 0.34 ± 0.09 | 122.7 ± 18.3 | | | |
| Bray | E1.4 | Inflamed | | 0.07 ± 0.08 | 1.91 ± 0.56 | 24.2 ± 28.5 | 56.3 | 0, 50, 100, 300, 600 | IVIM: bcNLLS ADC: weighted LLS |
| | | Normal | | 0.05 ± 0.04 | 0.92 ± 0.26 | 44.6 ± 35.2 | | | |

*Table 1 – Computational pipeline settings for E1. Synthetic signals were generated using parameters drawn from normal distributions taken from Zhao[38] or Bray[41]. Effective SNR for E1.4 was calculated from Bray's b=0 images, and, for Tasks E1.1-E1.3, adjusted by mean ROI size and acquisition differences (TE, voxel size, number of repetitions, etc.) between Bray and Zhao's experiments. In E1.4, ADC values were estimated from IVIM-synthesised data.*

### 3.3 – Benefits over current approaches (E2)

Two illustrative classification tasks were simulated to demonstrate the advantages our pipeline offers over current task-independent approaches to CED assessment.

Pipeline settings are shown in Table 2: for analytical clarity, tissue parameters were chosen such that only one parameter varied between subtypes per task. As in E1, these tasks involved classifying mean ROI parameter estimates as belonging to one of two SpA subtypes, and task performance was assessed with ROC curves and associated AUC. Data was synthesised from the IVIM model, with a single set of tissue parameters representing each SpA lesion subtype. IVIM parameter estimation was performed using both bcNLLS and sNLLS as above, with the exception that $b_{threshold}$ = 50 s/mm² within sNLLS fitting, a more appropriate choice than 200 s/mm² given E1.4's IVIM generative parameters.

Classification performance was assessed for different (a) models and (b) fitting methods. Any situation where the optimal choice of model or fitting method is found to be task-specific represents a failure of current task-agnostic CED approaches.

| Dataset | Task | Tissue | Generative model | $f$ | $D_{slow}$ ($10^{-3}\,mm^2/s$) | $D_{fast}$ ($10^{-3}\,mm^2/s$) | Effective SNR | Sampling ($s/mm^2$) | Model & fitting method |
|---|---|---|---|---|---|---|---|---|---|
| Simulated | E2.1 | Healthy | Intravoxel incoherent motion (IVIM) | 0.09 | 0.35 | 123 | 20 | 0, 10, 20, 40, 80, 100, 200, 400, 600 | IVIM: sNLLS & bcNLLS ADC: wLS |
| | | Chronic | | 0.12 | 0.35 | 123 | | | |
| | E2.2 | Active | | 0.12 | 0.60 | 123 | | | |
| | | Chronic | | 0.12 | 0.46 | 123 | | | |

*Table 2 – Computational pipeline settings for E2. Synthetic signals were generated from the IVIM model. Both IVIM and ADC parameters were estimated from IVIM-synthesised data.*

## 4 – RESULTS

### 4.1 – Validation (E1)

Figure 2 compares our Task E1.1-E1.3 predictions to ground-truth clinical ROC curves, and shows that our pipeline accurately predicts (a) the qualitative form of the ROC curves, (b) the relative task performance of different experimental settings, and (c) the absolute task performance (AUC) of each experimental design. These findings are replicated for Task E1.4 in Figure 3.

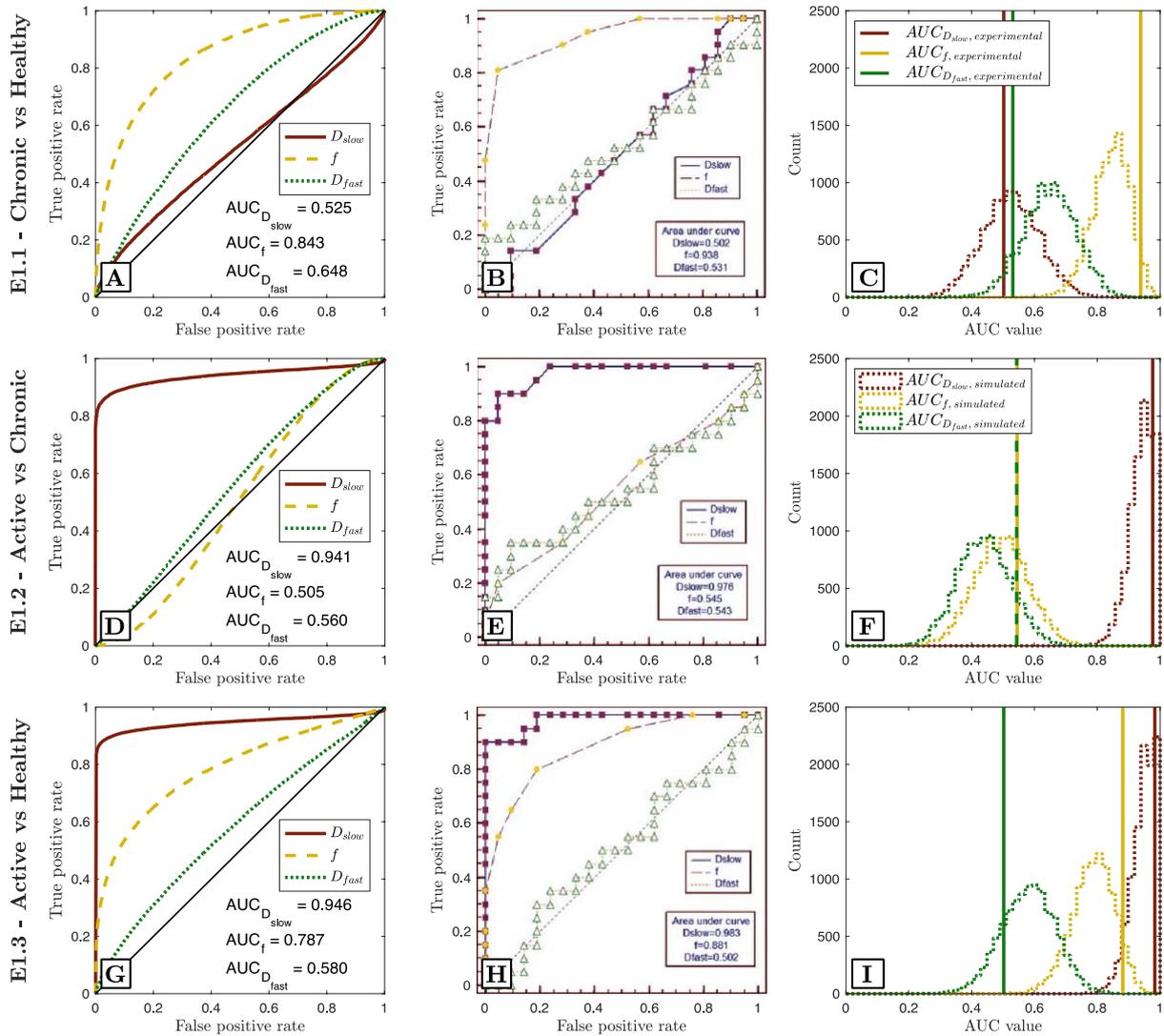

*Figure 2 – Simulated (first column) vs. clinical (middle column) ROC curves for Zhao's dataset (E1.1-E1.3). The third column compares clinical AUC values (vertical lines) to simulated AUC values (distributions) when sub-sampling simulated data to match clinical study sizes. All ROC curves are qualitatively similar; the relative performances (AUC values) of different IVIM parameters are equal; all AUC values are in numerical agreement once clinical sample size is considered.*

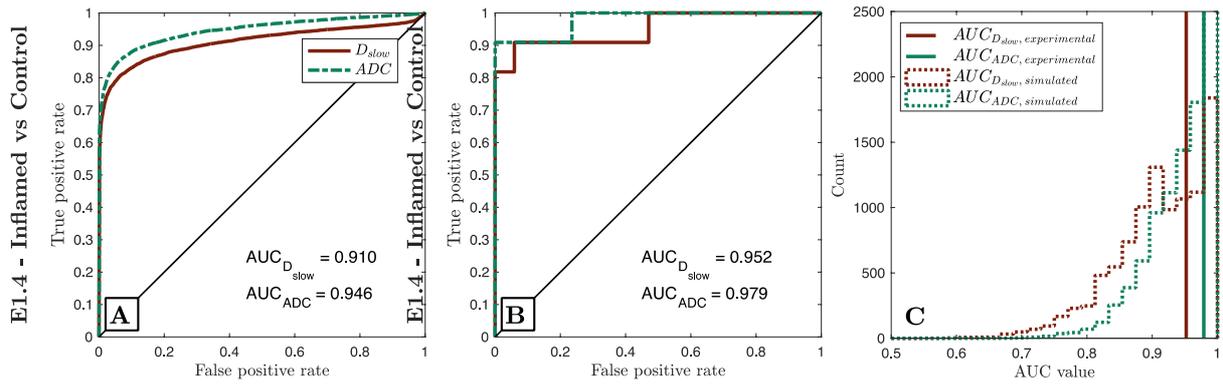

*Figure 3 – Simulated (first column) vs clinical (middle column) ROC curves for Bray's dataset (E1.4). The third column compares clinical AUC values (vertical lines) to simulated AUC values (distributions) when sub-sampling simulated data to match clinical study sizes. As in Figure 2, ROC curves are qualitatively similar; the relative performance (AUC values) of dMRI models are equal; all AUC values are in numerical agreement once clinical sample size is accounted for.*

## 4.2 – Benefits over current approaches (E2)

Figure 4 shows the advantages our pipeline offers over existing CED assessment methods. It demonstrates that, within a single disease, different classification tasks may be best served by different (a) models and (b) model fitting methods.

Regarding model selection, both ADC and IVIM models are sensitive to changes in microstructure (AUC > 0.5) in both E2.1 and E2.2. Within E2.1, where tissues differ by their perfusion fraction, ADC is less sensitive to population differences than IVIM ($AUC_{ADC}$ < $AUC_{IVIM}$, $CNR_{AUC}$ < $CNR_{IVIM}$). In contrast, in E2.2, where signal differences arise from variation in $D_{slow}$, the biased ADC model outperforms the ground truth generative model (IVIM). ADC, being the 'incorrect' model, gives upward-biased estimates of diffusivity, but results in improved classification performance ($AUC_{ADC}$ > $AUC_{IVIM}$), due primarily to its lower parameter estimation variance ($\sigma^2_{ADC} < \sigma^2_{IVIM}$). The upward bias does not adversely affect task performance as it is consistent across tissue types. These results demonstrate that (a) in general, biased models may outperform unbiased ones and (b) optimal model selection may vary, within a single tissue type, depending on the task of interest; model selection should not be task-agnostic.

With regard to parameter estimation, the optimal IVIM fitting method varies slightly between tasks (sNLLS for E2.1 and bcNLLS for E2.2). This result arises from the fact that fitting method performance varies across model parameters; the best method for estimating f (as needed in E2.1) may differ from that for estimating $D_{slow}$ (as per E2.2). Further data are included in supplementary materials, but these results again demonstrate that optimal experimental design choices are not task-agnostic.

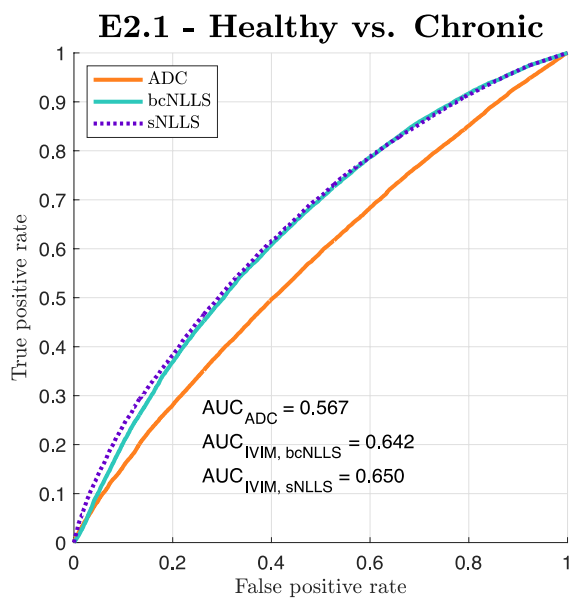
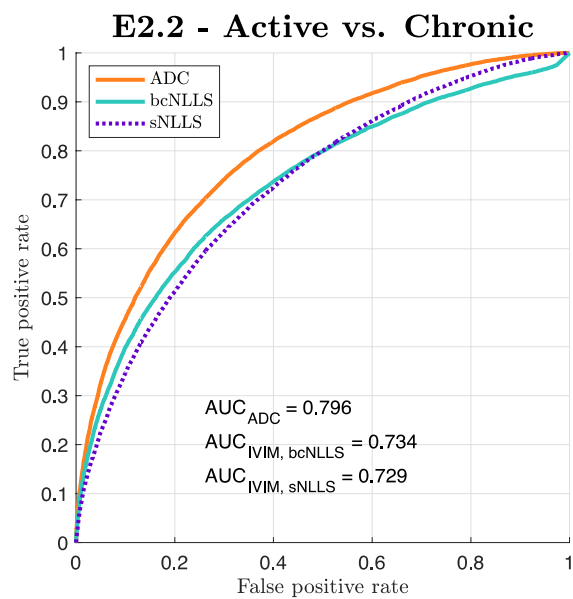
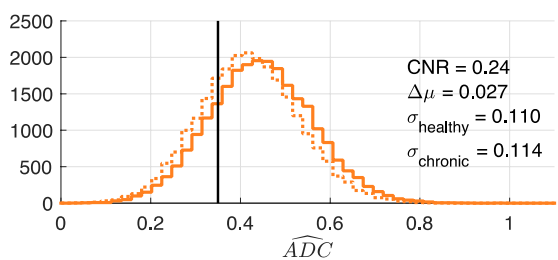
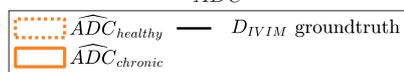
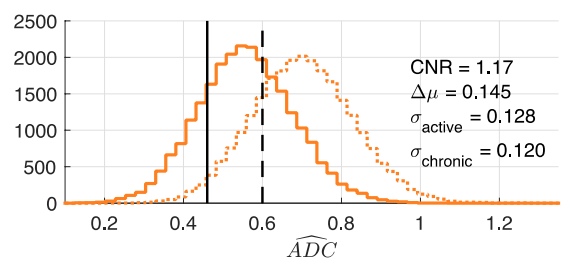
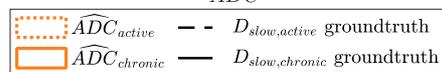
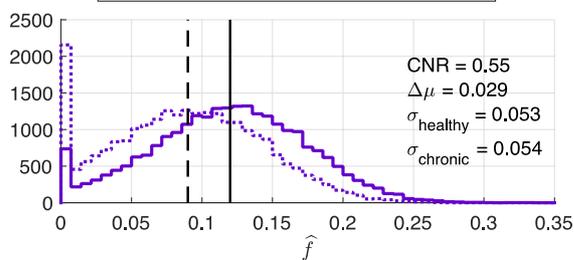
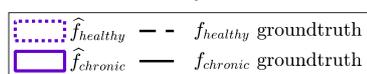
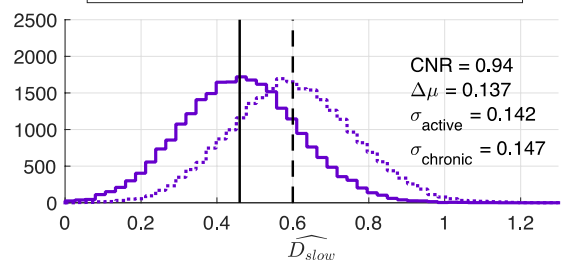
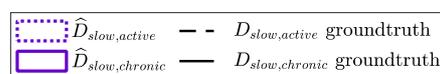

*Figure 4 – ROC curves and associated AUC values for two illustrative classification tasks, together with the distributions of parameter estimates underlying these ROC curves. Despite IVIM being the generative model for both tasks, it is outperformed by ADC in Task E2.2. Within IVIM, sNLLS outperforms bcNLLS in Task E2.1; the opposite is true in Task E2.2*

## 5 – DISCUSSION & CONCLUSIONS

This paper proposes and demonstrates a computational pipeline for assessing experimental designs on their task performance, rather than parameter estimation, capabilities. This mimics real-world task-driven experimental design, where clinical data is acquired and used to assess experimental settings.

Our method addresses limitations with current approaches to CED assessment, which rely on task-agnostic measures of parameter-estimation accuracy or precision, thereby providing indirect, potentially unreliable, predictions of task sensitivity and specificity. In contrast, by explicitly simulating the interactions between myriad experimental design choices, we are able to directly predict of task performance.

Validation experiments (E1) show that these predictions are accurate and reliable. This has three consequences. Firstly, our pipeline can be used to differentiate, between a set of candidate experimental settings, the single experimental design which maximises task performance. Secondly, our pipeline can be used to iteratively adjust experimental design parameters, such as acquisition time, until specific, clinically-required task performance is achieved. Thirdly, our pipeline's outputs can be used, with a high degree of confidence, to demonstrate the benefits our approach offers over current practice.

This demonstration, in the form of E2, shows that traditional parameter-estimation CED metrics lead to sub-optimal dMRI task performance. The fact that optimal model selection and fitting method varies across tasks, within pathology, is evidence of a failure of current practice. Assessed on parameter-estimation performance alone, IVIM beats ADC; yet, in Task E2.2, ADC yields higher task performance. Assessed on parameter-estimation performance, sNLLS beats bcNLLS[32,34]; yet, in Task E2.2, bcNLLS is the optimal IVIM parameter estimation algorithm. These findings demonstrate that task-agnostic parameter-estimation metrics, such as bias or variance, are unreliable predictors of task performance. Task-specific assessment, that directly measures experimental outcome, is required to reliably assess experimental designs. Such assessment is inaccessible within current CED practice.

**5.1 – Relation to existing work**

To the best of our knowledge, there is at present no similar CED framework for task-driven assessment. The closest to our approach is a method of MRI protocol optimisation driven by statistical decision theory[43]. The authors argue for a task-driven approach to protocol optimisation. However, the two approaches differ in an important way. Theirs maximises task performance via acquired MRI signals, which provide an indirect measure of the underlying difference in tissue properties. In contrast, our approach assesses task performance by directly considering the properties of interest: dMRI parameter estimates. Since tissue properties are estimated from measured MRI signals by means of model fitting, the choice of model and fitting methodology impacts task performance and is therefore explicitly assessed by our method.

Our work mimics, in-silico, the gold-standard method to assess task-driven experimental designs: acquiring rich, super-sampled clinical datasets which are successively sub-sampled, with each reduced dataset assessed on its associated task performance[32,35]. These methods are data-intensive and may be impractical for many applications. Our framework offers a computational alternative that 1) makes task-driven experimental assessment more accessible and 2) can be used to narrow the search space of experimental design choices before real data is required, thereby informing, focusing, and substantially shortening, any subsequent task-driven clinical evaluation and validation.

**5.2 – Use-cases and broader scope**

The proposed method produces assessment metrics for experimental designs, and it is left to the end-user to decide what to do with these metrics. The simplest use-case is experimental design selection: a range of plausible experimental settings are assessed, and the task-optimal experimental design is chosen from this set. Another use-case is optimisation, whether manual or automated: experimental settings are repeatedly adjusted and assessed; changes that improve task performance are retained, leading to iterative optimisation. Yet another use-case is calculating acquisition-time requirements: determining the acquisition protocol required for a specified task performance (e.g. <10% false-positive-rate); such calculations are not possible with existing task-agnostic assessment methods.

Regardless of use-case, the framework can be applied to a broad range of diffusion or quantitative MRI contexts. For simplicity, this work has focused on direction-averaged diffusion experiments, but the pipeline is compatible with any quantitative model-based task-driven application (e.g. fat fraction mapping[44]) for which a quantitative task (I1), well-characterised tissue properties (I2), and the ability to generate synthetic signal (I3) are available.

Furthermore, this work has used AUC as a summary task-performance metric due to the availability of clinical values to validate against. However, the framework's intermediate output – a distribution of parameter estimates underpinning ROC curves – can be used to construct a wide range of quality metrics based on specificity or sensitivity[45].

**5.3 – Limitations**

One potential limitation of the proposed approach is that it requires clear knowledge of the relationship between tissue diffusion properties and the underlying pathology of interest (i.e. I2). Whilst this insight may not currently exist for all clinical use-cases, it is a natural by-product of ongoing basic research; this work provides a computational framework to exploit these relationships.

Another potential limitation is the reliance of our approach on simulation. Simulation requires the use of computational models which inevitably represent an approximation of the underlying biophysical process. Nevertheless, the availability of advanced signal models (e.g. IVIM) which give good representations of measured dMRI data, allow our approach to inform experimental design choices in a wide range of clinical settings.

**5.4 – Outlook**

Our in-silico task-specific assessment promises improved experimental design selection and optimisation: all experimental choices, from data acquisition to analysis, can be analysed and compared without the need for expensive, time-consuming data acquisition. Model fitting methods can be assessed in a more meaningful, task-specific manner. Traditionally unfavoured, high-bias models, can be considered, assessed, and selected.

Although our pipeline can simply replace the assessment stage in current CED practice, it naturally lends itself to being incorporated into an overarching task-driven CED framework: combining the proposal of candidate experimental designs with computational optimisation, using the presented work as an accurate, task-specific optimisation metric.

**Supplementary material**

**Code availability**

Results and figures contained in this work can be reproduced by running the code found at: https://doi.org/10.5281/zenodo.5510276

**Task-specificity of optimal fitting method**

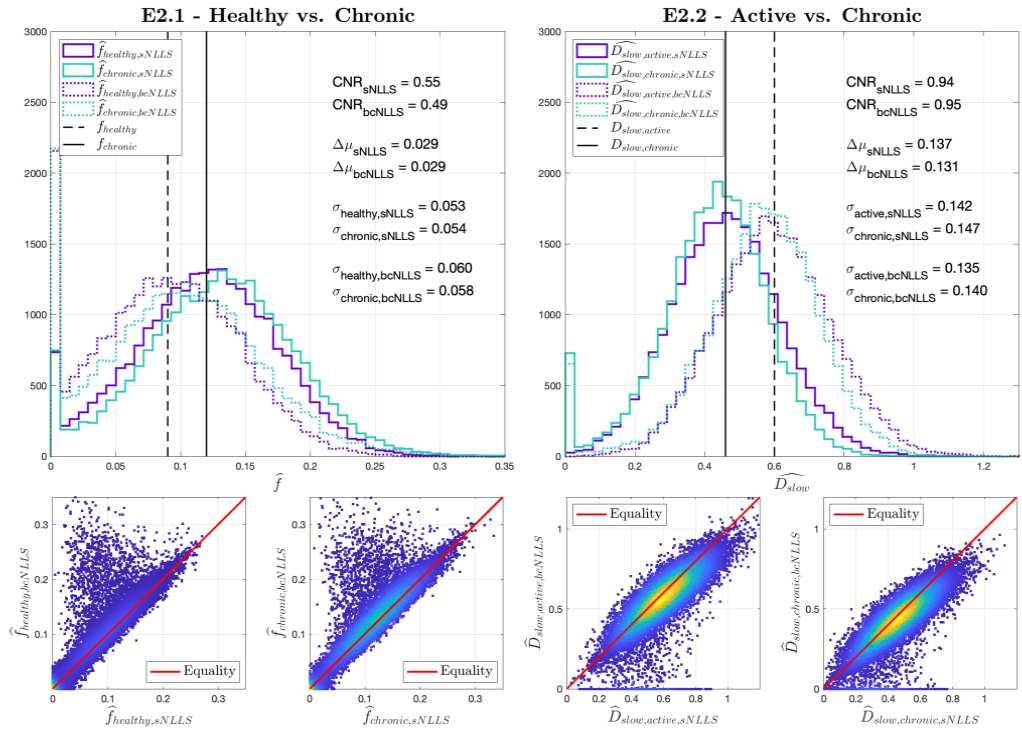

*Figure S1 – Comparison of sNLLS and bcNLLS fitting methods within tasks E2.1 and E2.2. The top panels show the distributions of IVIM parameter estimates; the bottom row shows the correlation between sNLLS and bcNLLS parameter estimates. The difference in fitting method performance is most pronounced in E2.1, where bcNLLS results in on-average higher $\hat{f}$ than sNLLS.*

# 6 – REFERENCES